\documentclass[prl,aps,showpacs,twocolumn]{revtex4}
\usepackage{amsmath}

\begin{document}
\title{Comment on `Spin Decoherence in Superconducting Atom Chips'}
\author{Stefan Scheel, E. A. Hinds, and P. L. Knight}
\affiliation{Quantum Optics and Laser Science, Blackett Laboratory,
Imperial College London, Prince Consort Road, London SW7 2BW, United
Kingdom}
\date{\today}
\begin{abstract}
We comment on a recent paper [Phys. Rev. Lett. \textbf{97}, 070401
(2006)] concerning rubidium atoms trapped near a superconducting
niobium surface at $\sim 4$\,K. This seeks to calculate the rate of
atomic spin flips induced by thermal magnetic noise. We point out
that the calculation is in error by a large factor because it is
based on the two-fluid model of superconductivity. This model gives
a poor description of electromagnetic dissipation just below the
critical temperature because it cannot incorporate the case II
coherences of a fuller quantum theory.
\end{abstract}


\maketitle

A recent publication \cite{PK} discusses atoms magnetically trapped
near a superconducting surface and attempts to calculate the spin
flip rate due to thermal magnetic noise. The Letter recounts the
formalism developed in \cite{Scheel04,Scheel05}, largely verbatim,
and then applies it to the two-fluid model of superconductivity.
This leads the authors to conclude that Rb atoms near a
superconducting niobium surface enjoy a lifetime increase of $10^5$
as the temperature is lowered by one or two degrees below $T_c\simeq
8$\,K. This disagrees dramatically with previous estimates of the
effect \cite{Scheel05,Henkel05} that were based on measured
electromagnetic dissipation in niobium. Here we point out that the
use of the two-fluid model has led to an erroneous result.

In the two-fluid model, the density of the normal component scales
as $(T/T_c)^4$. As the temperature drops, this rapidly suppresses
all dissipative processes, giving a very large derivative just below
$T_c$. This is precisely the behavior shown in figure 2 of
\cite{PK}. However, it is well known that the two-fluid model
provides a poor description of nuclear relaxation and
electromagnetic absorption in the temperature range just below
$T_c$. This is because of quantum interference effects in the
relevant low-energy scattering processes, which lead to the
so-called case II coherence factors. These cause the dissipation to
\textit{increase} as the temperature is lowered below $T_c$ before
it declines again in the low temperature limit. This behavior was
first observed in the case of nuclear-spin relaxation
\cite{HebelSlichter}, and is called the Hebel-Slichter peak. Chapter
3 of Tinkham's book \cite{Tinkham} gives a very accessible
introduction to this subject. There he points out that ``the ability
of BCS pairing theory, with its coherence factors, to explain this
difference [from the two-fluid model] in a natural way was one of
the key triumphs which validated the theory".

 More recently, the same type of behavior has been confirmed experimentally for
electromagnetic absorption  \cite{Klein94}, where the case II
coherence factors have been observed in superconducting Nb and Pb
over the range $T_c/4\lesssim T<T_c$. These are not pure BCS
superconductors but have to be described by the strong-coupling
Eliashberg theory. Nevertheless, their temperature-dependent
absorption exhibits the same type of peaked behavior, which
disagrees with the two-fluid model used in \cite{PK}.

The predictions for spin flip lifetimes made in previous papers
\cite{Scheel05,Henkel05} are based on experimental results which
explicitly confirm the existence of the case II coherence factor in
niobium \cite{Klein94}. The main enhancement of lifetime at $4$\,K
compared with room temperature ($\sim 100$) is due to the smaller
number of thermal photons per mode. In addition Ref.~\cite{Scheel05}
estimates that the lifetime should increase by a further factor of
$\sim 10$ as the temperature is lowered from $T_c$ to $T_c/2$, in
accordance with BCS theory and with the experimental evidence. By
contrast, Ref. \cite{PK} predicts an increase of $10^5$, not this
factor of 10, when cooling only by one or two degrees below $T_c$.
This is an artifact of the two-fluid model. Although the more
realistic calculation gives a smaller enhancement of the lifetime,
the effect is nevertheless exceedingly significant for the future of
superconducting atom chips \cite{Serge} as low-decoherence quantum
devices.


\end{document}